\documentclass[universe,article,accept,moreauthors,pdftex,10pt,a4paper]{Definitions/mdpi}
\usepackage{graphicx}
\usepackage{dcolumn}% Align table columns on decimal point
\usepackage{bm}
\usepackage{hyperref}% add hypertext capabilities
\firstpage{1}
\pubvolume{6}
\issuenum{11}
\articlenumber{206}
\pubyear{2020}
\copyrightyear{2020}
\history{Received: 14 October 2020; Accepted: 9 November 2020; Published: 11 November 2020}

\Title{Comparison between the Thomas--Fermi and Hartree--Fock--Bogoliubov Methods in the Inner Crust of a Neutron Star: The Role of Pairing Correlations%Attention AE/ME. The following layout issues have not been checked by the English Editing Department and must be carefully verified by the AE/Layout Department: All callout issues, bold usage of callouts, and references to callouts in the text. Correct callout usage in figures. Figure and Table layout issues. Footnote formatting and Glossaries have not been checked. En dash usage for negative values, en dash usage to indicate relationships, en dash usage to indicate bonds (especially in chemistry). The English Editing Department is not responsible for correct italic usage for genes, proteins and technical terminology. This responsibility belongs to the authors. The following are also not checked: spacing between numbers and units of measurement, ratios, en dashes for ranges, date and time formats, punctuation in equation lines, and less than/more than spacing (< >). Finally, capitalization and layout of titles/headings must be properly checked and ensuring 'Eq.' and 'Figure ' are properly spelled out, as these are layout issues.
}
\Author{Matthew Shelley *\href{https://orcid.org/0000-0002-7792-3618}{\orcidicon} and Alessandro Pastore\href{https://orcid.org/0000-0003-3354-6432}{\orcidicon}}
\AuthorNames{Matthew Shelley and Alessandro Pastore}
\address[1]{%
Department of Physics, University of York, %Shelley: Added %MDPI: Please add the department/school/faculty/campus.
Heslington, York YO10 5DD, UK; alessandro.pastore@york.ac.uk
}
\corres{\hangafter=1 \hangindent=1.05em \hspace{-0.82em} Correspondence: mges501@york.ac.uk}
\abstract{We investigated the role of a pairing correlation in the chemical composition of the inner crust of a neutron star with the extended Thomas--Fermi method, using the Strutinsky integral correction. We compare our results with the fully self-consistent Hartree--Fock--Bogoliubov approach, showing that the resulting discrepancy, apart from the very low density region, is compatible with the typical accuracy we can achieve with standard mean-field methods.}
 %Shelley: Added %MDPI: please add keywords.
\begin{document}
\section{Introduction}\label{sec:intro}

The quest to find the equation of state~(EoS) that best describes the properties of neutron stars (NS)~\cite{chamel2008physics} is one of the major challenges in nuclear physics. NSs are extremely compact objects, and~so the density and pressure through their interior spans several orders of magnitude, and~consequently it is important to use a theoretical model that can accurately describe such a large density~range.

The tool of choice to describe both finite nuclei and NS is the nuclear energy density functional~(NEDF)~\cite{Book:Reinhard2004}. By~carefully adjusting the parameters of the functional~\cite{Goriely2009}, it is possible to obtain a {unified} equation of state~\cite{pearsonUnifiedEquationsState2018} that can describe the entire NS, from~the low-density outer region to the core.
Thanks to the latest advances both in the way one observes them~\cite{gendreau2012neutron} and the technique used~\cite{abbott2017gw170817}, it is possible to provide additional constraints to the EoS~\cite{most2018new,blaschkePhasesDenseMatter2018}. By~combining those with more traditional constraints based on heavy-ion collision experiments~\cite{stoecker1986high,danielewicz2002determination} it is possible to obtain interesting information about the properties of nuclear matter at high densities. By~combining all this information, together with the most recent measurements of finite nuclei~\cite{Steiner2008}, it is possible to construct accurate models to describe the physics of such massive~objects.

Due to the extreme pressure gradient, NS matter is not homogeneous. With~current models~\cite{chamel2008physics}, one can identify two main regions: the {crust}, which has a crystalline structure, and~the {core}, which is in a liquid phase whose composition is still under debate~\cite{lattimer2004physics, alford2007quark, chatterjee2016hyperons, vidana2018d, li2018competition}. The~crust spans a density range of $\approx10^{-8}\rho_0$ to $\approx0.5\rho_0$, where $\rho_0=0.16$~fm$^{-3}$ is the nuclear saturation~density.

The crust can be further divided into two regions: the {outer} crust and the {inner} crust. From~the earliest models~\cite{baym1971ground} to more modern ones~\cite{pearsonPropertiesOuterCrust2011,chamelBinaryTernaryIonic2016}, it has been predicted that the outer crust has a crystalline structure of finite nuclei, surrounded by a gas of free electrons. The~inner crust has a similar structure, but~because of the higher density, neutrons start to drip out from the nuclei and form a gas~\cite{negele1973neutron, chamel2015neutron,pastore2020impact}.

Modeling the inner crust structure is particularly challenging, since it requires the treatment of the clusters and gas on an equal footing. Ideally one should use band theory, as~typically used in solid state physics, to~discus the properties of crystal with delocalized electrons~\cite{chamel2007validity, chamel2012neutron}.

A widely accepted simplification consists of defining a spherical Wigner--Seitz~(WS) cell centered on each cluster at a given baryonic density, and~to assume no interaction between cells. The~size of the cell is determined by assuming charge neutrality, and~that protons, neutrons and electrons are at $\beta$-equilibrium within the WS cell. We refer to reference~\cite{chamel2007validity} for a detailed discussion on the validity of such approximation.
Having defined the main features of the system, one can then determine the proton number~$Z$ within the cell, and~the cell radius~$R_{WS}$, after~minimizing the total energy per particle~$e$ of the system. This then yields the total particle number~$A$, and~the proton fraction~$Y_p=Z/A$, a~quantity of astrophysical~importance.

Within the NEDF framework, one should calculate the nuclear contribution to the total energy of the WS cell using the Hartree--Fock--Bogoliubov~(HFB)~\cite{ring2004nuclear} equations, since they provide a fully quantum mechanical description of the system without making an artificial distinction between bound and unbound neutrons. The~downsides of the HFB method are its computational cost and the possible numerical inaccuracies related to the boundary conditions adopted in the calculation~\cite{baldo2006role,Margueron2007b,pastore2011superfluid}. %Shelley Removed double reference (27-28)
We refer to reference~\cite{pastore2017new} for a more detailed discussion. More recently, the~HFB method has been used to investigate heating of the crust~\cite{fantinaCrustalHeatingAccreting2018} and the crust-core transition~\cite{schuetrumpfSurveyNuclearPasta2019}, though~these are not systematic studies of the~crust.

To overcome the numerical difficulties that arise in the HFB method, but~also to reduce the computational cost, several authors have adopted semiclassical methods based on the extended Thomas--Fermi~(ETF)~\cite{brack1976extended} approximation. To~account for nuclear shell structure, an~energy correction is added using the Strutinsky integral~(SI) method~\cite{pearsonInnerCrustNeutron2012}. Combined, these are named the ETFSI~method.

The results obtained using the two methods seem to qualitatively disagree. Using HFB, one~typically observes clusters with a variety of different $Z$ through the crust~\cite{grill2011cluster,pastore2017new}. On~the other hand, ETFSI points towards an inner crust comprising clusters with only $Z=40$, both with Skyrme forces~\cite{pearsonInnerCrustNeutron2012,pearson2015role,pearsonUnifiedEquationsState2018} and with finite-range Gogny forces~\cite{mondalStructureCompositionInner2020}.
% The reason for this disagreement could be related to the properties of the functional used to perform the calculations, or some aspect of the approximations in the ETFSI method.
The main reason for this disagreement is probably related to the properties of the functional used to perform the calculations. In~previous work~\cite{shelley2020accurately}, we compared the structure of some WS cells performing both ETFSI and HFB calculations using the same functional. We observed an energy discrepancy far larger than the estimated error of the HFB method~\cite{pastore2017new}, and~we identified the cause as a lack of neutron pairing correlations~\cite{pastore2013pairing} in the ETFSI approach~\cite{pearson2015role}.
% For this purpose, a preliminary investigation was carried out in reference~\cite{shelley2020accurately}: by comparing ETFSI and HFB calculations using the same functional, and performing the calculations in exactly the same WS cell, the lack of neutron pairing correlations~\cite{pastore2013pairing} in the ETFSI approach~\cite{pearson2015role} was identified as a possible source of the energy discrepancy.

In the present article, we therefore present a simple methodology to incorporate the neutron pairing energy contribution into the ETFSI method, and~we compare the results obtained with those from HFB~calculations.

The article is organized as follows: in Section~\ref{sec:ws} we briefly discuss the energy contributions within a WS cell, and~we also illustrate the ETFSI method with the extension to include pairing correlations. We present our main results for the inner crust structure in Section~\ref{sec:results}, and~we provide our conclusions in Section~\ref{sec:concl}.

%%%%%%%%%%%%%%%%%%%%%%%%%%%%%%%%%%%%%%%%%%%%%%%%%%%%%%%%%%%%%%%%%%%%%%%%%%%%%%%%%%%%%%%%%%%%%%%%%%%%

\section{The Wigner--Seitz~Cell}\label{sec:ws}

Adopting the Wigner--Seitz~(WS) approximation, we divide the periodic lattice of nuclei of the inner crust into spherical charge-neutral non-interacting cells. For~a given baryonic density~$\rho_b$, all cells are assumed identical. Each contains a nuclear cluster at its center surrounded by a gas of superfluid neutrons, and~also contains a near-homogeneous ultra-relativistic electron gas.
The radius of a cell $R_{WS}$ is defined as half of the distance between neighboring clusters. Under~the condition of charge neutrality, the~number of protons $Z$ in the WS cell must equal the number of electrons. For~given values of $\rho_b$ and $R_{WS}$, the~total number of particles $A$ is fixed by the relation
\begin{equation}\label{eqn:tot_particles}
A=\rho_b \frac{4\pi}{3}R_{WS}^3\;.
\end{equation}

Consequently, a~WS cell is uniquely defined by the parameters \{$\rho_b,Z,R_{WS}$\}. At~zero temperature, the~total energy per particle of the system is given by
\begin{equation}\label{eqn:etot}
e = e_{Sky} + e_e - Y_p Q_{n,\beta}\;,
\end{equation}
where $e_{Sky}$ is the contribution arising from the baryons interacting via the strong force and from the Coulomb interaction between protons, while $e_e$ is sum of the kinetic and potential energies of ultra-relativistic electrons~\cite{shapiro2008black} and the proton-electron interaction~\cite{grill2011cluster}. The~last term accounts for the mass difference between neutrons and protons, $Q_{n,\beta}=0.782$ MeV. $Y_p=Z/A$ is the proton fraction of the~cell.

The terms related to electrons ($e_e$) and to the Coulomb component of $e_{Sky}$ are treated on equal footing for the HFB and ETFSI methods, as~described in~\cite{pearsonInnerCrustNeutron2012}. The~ energy contribution per particle for a Skyrme-type nucleon-nucleon interaction~\cite{skyrme1956cvii} is expressed as functional of local densities
\begin{eqnarray}
e_{Sky}=\frac{1}{A}\int \mathcal{E}\left(\rho_q(\mathbf{r}),\tau_q(\mathbf{r}),\vec{J}(\mathbf{r}) \right)d\mathbf{r}\;,
\end{eqnarray}
where $q=n,p$ stands for the nuclear charge. Considering only time-reversal invariant systems, the~Skyrme functional depends only on a linear combination of the matter densities $\rho_q(r)$, kinetic~densities $\tau_q(r)$, the~spin current densities $\vec{J}(r)$ and their derivatives~\cite{perlinska2004local}.
The procedure used to calculate the local densities differs in the HFB and semiclassical methods, and~it is outlined in the \mbox{Sections~\ref{subsec:HFB} and~\ref{subsec:etf}}.

\subsection{Hartree--Fock--Bogoliubov}\label{subsec:HFB}

In the HFB approach, the~densities (and thus the fields) are calculated directly using the quasi-particle wave-functions, which are the solutions of the HFB equations~\cite{ring2004nuclear}.
\begin{eqnarray}\label{eqn:HFB}
\sum_{n'} (h^q_{n'nlj}-\varepsilon_{F,q})U_{n'lj}^{iq}+\sum_{n'}\Delta^q_{nn'lj}V^{iq}_{n'lj}=E^q_{ilj}U^{iq}_{nlj}\\
\sum_{n'}\Delta^q_{nn'lj}U^{iq}_{n'lj}-\sum_{n'} (h^q_{n'nlj}-\varepsilon_{F,q})V_{n'lj}^{iq}=E^q_{ilj}V^{iq}_{nlj}
\end{eqnarray}
where $\varepsilon_{F,q}$ is the Fermi energy. We used the standard notation~$nlj$ for the spherical single-particle states with radial quantum number~$n$, orbital angular momentum~$l$ and total angular momentum~$j$. $V_{nlj}^{iq}$~and $U_{nlj}^{iq}$~are the Bogoliubov amplitudes for the $i$-th quasi-particle with energy~$E^q_{ilj}$. The~single-particle Hamiltonian~$h$ is built from the Skyrme functional, while $\Delta_{nn'lj}^q$ are the matrix elements of the pairing gap obtained from a contact pairing interaction. In~the case of vanishing pairing, these equations reduce to the Hartree--Fock~(HF) one.
For more details on the numerical methods used to solve these equations we refer to references~\cite{pastore2017new,pastore2011superfluid,pastore2012superfluid,pastore2013pairing}.
The most relevant point for the following discussion is the choice of the boundary conditions used to solve Equation~\eqref{eqn:HFB}.
In the present article, we use the Dirichlet--Neumann mixed boundary conditions: (i)~even-parity wave-functions vanish at $r = R_{WS}$; (ii)~the first derivative of odd-parity wave~functions vanishes at $r = R_{WS}$. We call them Boundary Conditions Even (BCE), in~contrast to the boundary conditions odd (BCO) where the two parity states are treated in the opposite way.
We have checked that this particular choice does not affect the final results. See also reference~\cite{pastore2017new} for more~details.

\subsection{The ETFSI~Method}\label{subsec:etf}

In this section, we briefly describe the extended Thomas--Fermi + Strutinsky integral method (ETFSI), as~developed in references~\cite{Grammaticos1979,bartel2002nuclear,onsi2008semi,pearsonInnerCrustNeutron2012}.
Within the semiclassical approach, the~neutron and proton densities are parameterized. Assuming no proton gas, we use a generalized Fermi--Dirac distribution of the form
\begin{eqnarray}\label{eqn:fd_profile}
\rho_q(r)=\frac{\rho_q^{liq}-\delta_{q,n}\rho^{gas}}{1+\exp\left(\frac{r-r_q}{a_q} \right)}+\delta_{q,n}\rho^{gas}\;.
\end{eqnarray}

$\rho_q^{liq}$ are the densities of the neutrons and protons at the center of the WS cell, $r=0$, while $\rho^{gas}$ is the density of neutrons at the edge, $r=R_{WS}$. $r_q$ are the cluster radii of the neutrons and protons, and~$a_q$ are the diffusivities of the cluster surface.
These 7 adjustable parameters are determined by the minimization of the energy per particle, given in Equation~\eqref{eqn:etot}, under~the constraints of charge neutrality and $\beta$-equilibrium. See Section~\ref{subsubsec:min} for more~details.

The authors of references~\cite{onsi2008semi,pearsonInnerCrustNeutron2012,pearsonUnifiedEquationsState2018} have introduced an additional damping factor in Equation~\eqref{eqn:fd_profile}. Although~such a term may be useful for avoiding convergence problems at high $\rho_b$, it has a small impact on the energy per particle of the system, and~for the following analysis we can safely proceed without~it.

The kinetic and spin-current densities are expressed as a function of the matter density $\rho_q$ and its derivatives via the Wigner--Kirkwood expansion~\cite{ring2004nuclear}. In~the present work, we use the full expansion, up~to 4th order in gradients, employing the explicit expressions for the 4th-order density contributions as given in the appendix of reference~\cite{bartel2002nuclear}. This differs from the approach of reference~\cite{pearsonInnerCrustNeutron2012}, explained~in Section II of this~reference.

\begin{figure}[H]
\centering
\scalebox{.85}[.85]{\includegraphics{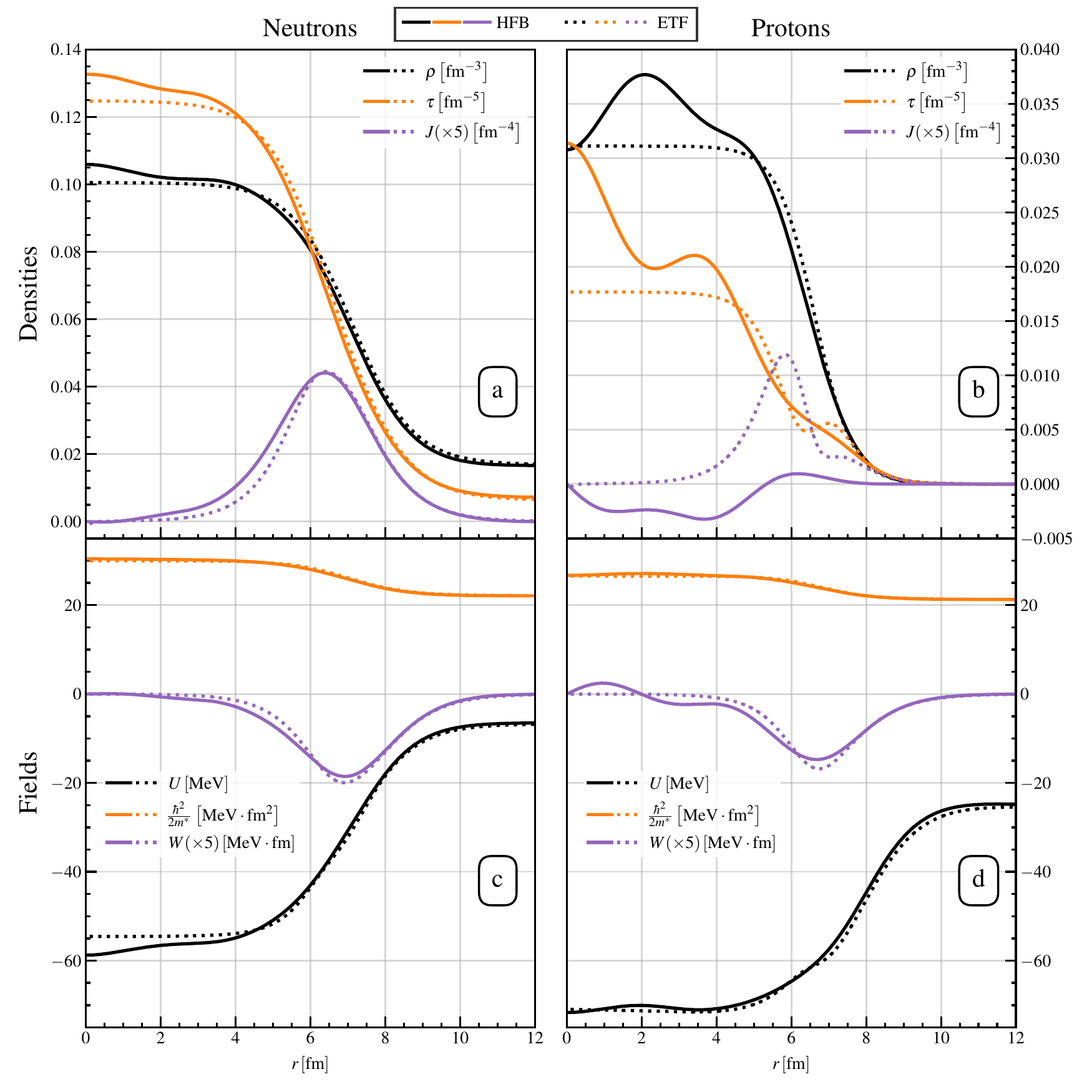}}
\caption{Densities %Shelley: (a)--(d) are used for reference in the text %MDPI: There is no explanation for (a)--(d) in the figure.
(top row) and fields (bottom row), for~neutrons (\textbf{left column}) and protons (\textbf{right~column}), for~a Wigner--Seitz cell with $\rho_b=0.02$~fm$^{-3}$, $Z=40$ and $R_{WS}=24.0$~fm. Black~lines show the matter densities and central potentials; orange lines show the kinetic densities and effective masses; and purple lines show the spin current densities and spin-orbit fields. Solid lines show the results of using the HFB method, and~dotted lines show those from using the ETFSI method; both used the SLy4 functional. Note the different scale used in panels \emph{a} and \emph{b} for the neutron and proton densities, to~make clearer the bump observed in the proton densities; see the text for details. Note also that all spin current densities and spin-orbit fields are multiplied by 5 to make them more~visible.}
\label{fig:densities_fields}
\end{figure}

We now examine the quality of our semiclassical method by comparing the densities and fields with those obtained from a fully self-consistent HFB calculation.
In Figure~\ref{fig:densities_fields}, we illustrate the density profiles for a WS cell with $\rho_b=0.02$~fm$^{-3}$, $Z=40$ and $R_{WS}=24.0$~fm, obtained using the SLy4 functional~\cite{chabanat1998skyrme}, by~solving the HFB equations. In~the same figure, we also illustrate the results obtained using ETFSI: the seven parameters characterizing the semiclassical matter densities~(Equation~\eqref{eqn:fd_profile}) were adjusted to reproduce the matter densities obtained from the HFB~calculation.

We observe that the 4th order expansion works nicely, and~reproduces very well the neutron kinetic and spin-current densities, as~shown in Figure~\ref{fig:densities_fields} (panel~\emph{a}). The~proton kinetic density presents a small {bump} around $7$~fm, as~seen in panel~\emph{b} of Figure~\ref{fig:densities_fields}. Due to the large density gradient in the proton cluster surface compared to the neutron one,  the~4th order truncation is probably not fully satisfactory. The~main consequence is the poor reproduction of the proton spin current density. This is not a major issue since the spin-orbit field is not too affected by such a~difference.

From the densities, one obtains the corresponding fields~\cite{perlinska2004local}: the central potentials $U_q(r)$, the~effective masses $\frac{\hbar^2}{2m_q^*(r)}$ and the spin-orbit fields $W_q(r)$. They are shown in the lower panels of Figure~\ref{fig:densities_fields}, together with the corresponding ones obtained solving the HF equations. In~both neutron and proton cases the agreement is very good, thereby showing the validity of the semiclassical approximation in capturing the main features of the HFB calculation. The~{bump} in the kinetic and spin current proton densities translates to an oscillation of the central proton potential at $r\approx7$~fm, but~the impact is~small.

Under the ETF approximation, shell effects are not accounted for. Consequently, the~authors of reference~\cite{onsi2008semi} suggested including (at least for protons) a perturbative contribution to the total energy using the Strutinsky integral~(SI) theorem, without~acting on the densities or fields. For~the sake of completeness we implemented exactly the same~method.
\unskip

\subsection{Choice of~Functionals}\label{subsec:functionals}

\begin{figure}[H]
\centering
\includegraphics{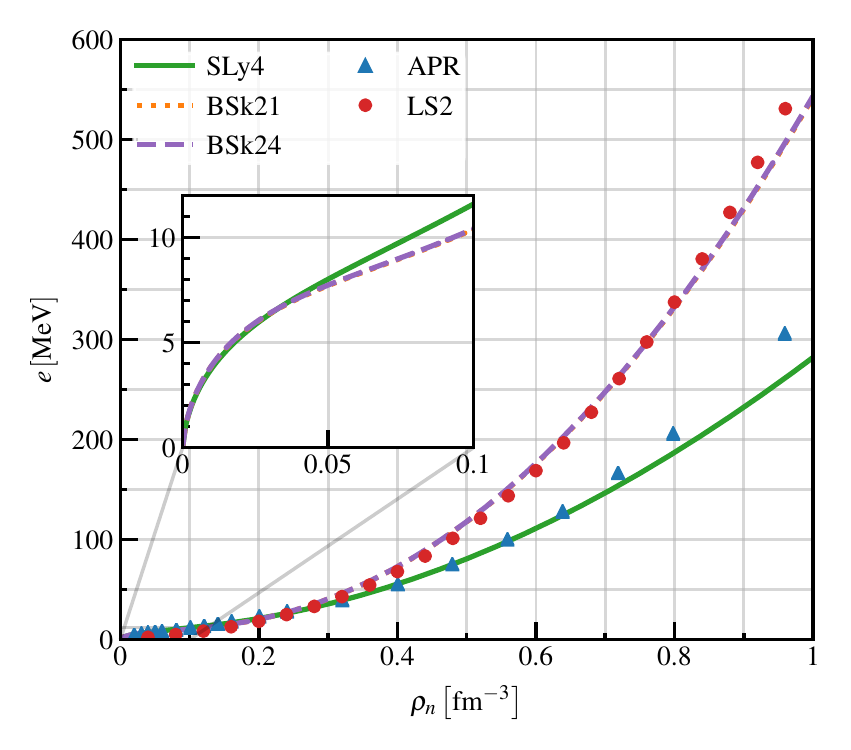}
\caption{Energy per particle $e$ in pure neutron matter for the three functionals used. The~EoS calculated in references~\cite{wiringaEquationStateDense1988,wiringaDeuteronsNeutronStars1993}~(APR), and~the one calculated in reference~\cite{liNeutronStarStructure2008}~(LS2) are also~shown.}
\label{fig:PNM_rho}
\end{figure}

Before more detailed analysis on the results obtained with HFB and ETFSI, we briefly discuss the choice of the Skyrme functional used in this work.
In Figure~\ref{fig:PNM_rho} we show the energy per particle $e$ curves for pure neutron matter (PNM) as a function of the density of the system for the three functionals we investigated: SLy4~\cite{chabanat1998skyrme}, BSk21~\cite{pearsonInnerCrustNeutron2012} and BSk24~\cite{goriely2013further}. Their parameters were all adjusted for the functionals to be applicable to both finite nuclei and higher density neutron star matter, for~modeling an entire neutron star.
On the same we also show the EoS calculated using {ab~initio} methods and given in references~\cite{wiringaEquationStateDense1988,wiringaDeuteronsNeutronStars1993}~(APR), and~the one calculated in reference~\cite{liNeutronStarStructure2008}~(LS2).
Both BSk21 and BSk24 have been fit on the LS2 EoS, while SLy4 has been adjusted using~APR.

The SLy4 functional was fit to doubly-magic nuclei; as a consequence, we are left with some freedom in choosing how to model pairing correlations.
In this work, we use a simple density-dependent pairing interaction of the form~\cite{bertsch1991pair}
\begin{eqnarray}\label{eq:pair}
v^{pair}_q(\mathbf{r}_1,\mathbf{r}_2)=v_{0q}\left[1-\eta \left(\frac{\rho_q(r)}{\rho_0}\right)^\alpha\right]\delta\left(\mathbf{r}_1-\mathbf{r}_2\right)\;.
\end{eqnarray}

\begin{figure}[H]
\centering
\includegraphics{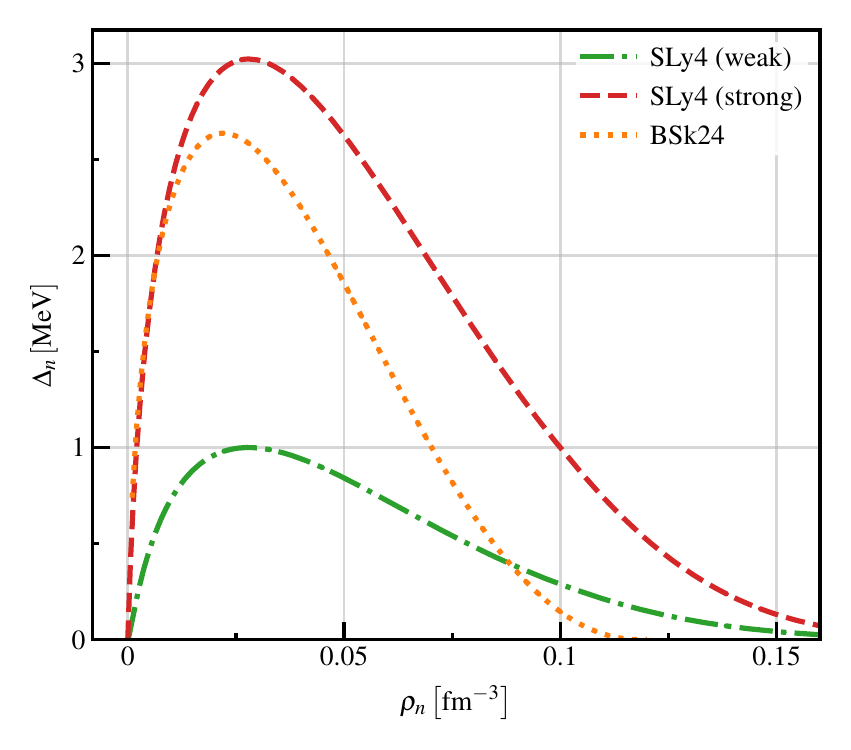}
\caption{The pairing gap in pure neutron matter (PNM), as~a function of density, for~the two pairing strengths used with the SLy4 functional (green dash-dotted line shows weak, red dashed line shows strong), and~for the BSk24 functional (orange dotted line). See text for~details.}
\label{fig:PNM_gaps}
\end{figure}

We choose the parameters $\eta=0.7$ and $\alpha=0.45$. We assume that the pairing strength is the same for neutrons and protons and we fix the pairing strength $v_{0q}$ to obtain a maximum pairing gap in PNM of $\approx$3~MeV, hereafter named {strong}, or~a maximum of $\approx$1~MeV, hereafter named {weak}, as~done in reference~\cite{grill2011cluster}. These choices largely cover the available range of results concerning the density evolution of the pairing gap in infinite nuclear matter~\cite{gandolfi2008equation}.
To avoid the ultraviolet divergence of the interaction given in Equation~\eqref{eq:pair} ~\cite{bulgac2002renormalization}, we adopt a smooth cut-off in quasi-particle space at $E_{ijl}^q\ge20$~MeV that is defined by an Gaussian factor $\exp\left(\left(E_{ijl}^q-20\right)/100\right)$.
In Figure~\ref{fig:PNM_gaps}, we report the density dependence of the pairing gap in PNM obtained by solving the BCS %Define, if~appropriate.
equations. The~maxima of these pairing gaps are located at $\rho_n\approx0.03$~fm$^{-3}$~(SLy4), corresponding to a Fermi momentum of $k_F^n\approx0.94$~fm$^{-1}$.

The pairing interactions for the BSk21-24 functionals~\cite{chamelEffectiveContactPairing2010} have been adjusted to reproduce the $^1S_0$ pairing gaps in both symmetric nuclear matter and PNM, as~obtained from Brueckner calculations using the Argonne~$V18$ nucleon--nucleon potential~\cite{cao2006screening}.
The resulting pairing gap is also reported in Figure~\ref{fig:PNM_gaps} as a function of the neutron density. We observe that in this case, the~pairing gap reaches a maximum of $\Delta_n\approx2.5$~MeV around $\rho_n\approx0.02$~fm$^{-3}$, corresponding to a Fermi momentum of $k_F^n\approx0.87$~fm$^{-1}$.

Within the literature, there is a wide consensus on the importance of pairing correlations within the inner crust of neutron stars~\cite{barranco1997role,dean2003pairing,baldo2006role,grill2011cluster,pastore2011superfluid,sandulescu2004superfluid,chamel2012neutron,maurizio2014nuclear,watanabe2017superfluid,bennemann2008superconductivity}.
While pairing correlations are naturally included within the HFB equations, Equation~\eqref{eqn:HFB}, the~original ETFSI method~\cite{onsi2008semi} is not capable of treating such correlations. As~a consequence, in~reference~\cite{pearson2015role} the authors have modified the ETFSI formalism to include the effects of proton pairing with the BCS approximation~\cite{ring2004nuclear}, but~still with no explicit treatment of neutron pairing~correlations.

Since the neutrons in the inner crust form a gas, it is not possible to apply directly the same methodology without running into the same type of problems encountered with the HFB method, which are related to spurious shell effects in the neutron gas~\cite{Margueron2007b}.
In this work, we have developed an additional energy correction based on the local density approximation (LDA).
A similar approach was already proposed in reference~\cite{burrelloHeatCapacityNeutron2015}. %Shelley: Reference added at end of file %Please confirm this ref citation.
In the the weak-coupling limit, the~correction to the energy per particle from superfluid neutrons is~\cite{PhysRevLett793347}
\begin{eqnarray}\label{eqn:lda}
e_{cond}=-\frac{3\Delta_n^2}{8\mu_n}\;.
\end{eqnarray}

The chemical potential $\mu_n$ is approximated by the corresponding Fermi energy $\varepsilon_{F,n}=k_{F,n}^2\hbar^2/2m^*_n$, where $k_{F,n}=\left(3\pi^2\rho_n\right)^{1/3}$. $\Delta_n=\Delta_n(\rho_n(r))$ is the local pairing gap as extracted from PNM calculations at a given density $\rho_n$ as discussed in reference~\cite{pastore2008microscopic} and illustrated in Figure~\ref{fig:PNM_gaps}.
We refer to Appendix~\ref{app:cond} for more details. The~energy correction in Equation~\eqref{eqn:lda} can be easily implemented in the ETFSI formalism without a major increase in the computational~cost.

\begin{figure}[H]
\centering
\scalebox{.85}[.85]{\includegraphics{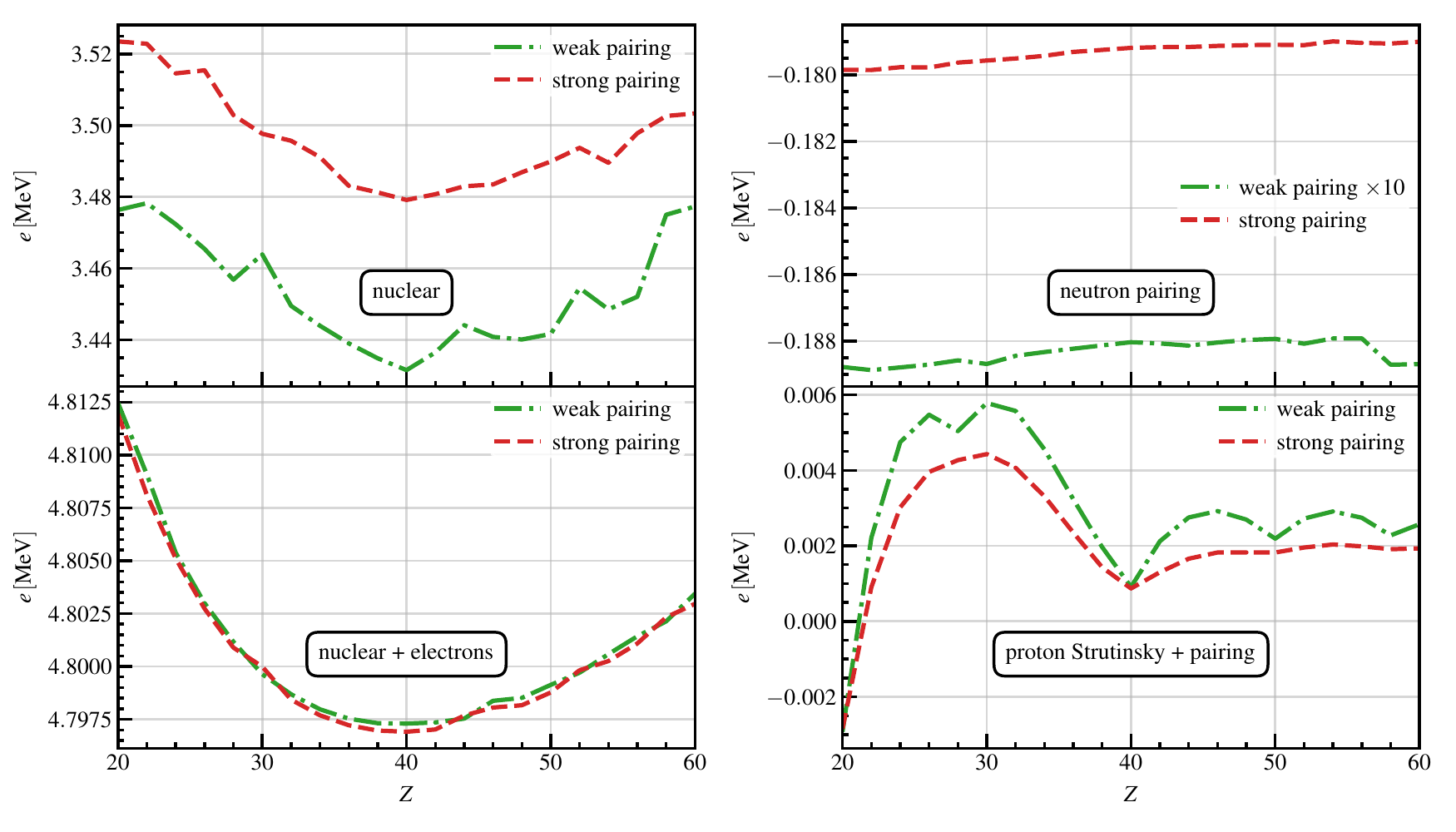}}
\caption{Individual contributions to energy per particle, $e$, for~a Wigner Seitz cell with $\rho_b=0.02$~fm$^{-3}$. Results are shown for the two pairing strengths used with the SLy4 functional (green dash-dotted line shows weak, red dashed line shows strong). The~top left panel shows only $e_{Sky}$, while the bottom left shows $e_{Sky}+e_e$; see Equation~\eqref{eqn:etot} and text for details. In~the top-right panel for neutron pairing, showing $e_{cond}$ (Equation~\eqref{eqn:lda}), the~weak result has been multiplied by 10 to make clear the variations for weak and strong on the same~scale.}
\label{fig:e_contributions}
\end{figure}

In Figure~\ref{fig:e_contributions} we illustrate the different contributions to the energy per particle $e$ for various WS cell at fixed baryonic density $\rho_b=0.02$~fm$^{-3}$, but~for the cases of {weak} and {strong} pairing. The~ETFSI+pairing results have been obtained using a complete minimization of the total energy of the WS cell using the SLy4 functional.
In the upper-left panel we show the {nuclear} contribution, referring~to the ETF energy including proton-proton Coulomb interaction. The~higher $e$ with stronger pairing simply reflects the larger neutron number obtained. The~higher the proportion of neutrons in the system, the~more neutron pairing can occur, lowering the total energy. When we include also the electron-electron and electron-proton interactions, as~shown in the lower-left panel, we see almost no difference between the {weak} and {strong} pairing cases. For~{strong} pairing, the~increase in the nuclear energy is offset by a decrease in the energy contribution from the electrons, a~result of the larger WS~cell.

In the upper-right panel of Figure~\ref{fig:e_contributions}, we see that the contribution to $e$ from neutron pairing is almost flat with respect to $Z$, for~{weak} and {strong}. Note that the {weak} case has been multiplied by 10 in this panel. In~Figure~\ref{fig:PNM_gaps}, where the PNM gap is about three times higher for the {strong} interaction at $\rho_b=0.02$~fm$^{-3}$, and~Equation~\eqref{eqn:lda}, showing the quadratic dependence of $e_{cond}$ on the PNM gap, the~factor of nine increase in the neutron condensation energy was~expected.

The energy per particle coming from the SI correction and proton pairing energy, as~calculated in reference~\cite{pearson2015role}, are shown in the lower-right panel of Figure~\ref{fig:e_contributions}. The~shell effects give rise to local minima at $Z=20,28,40,50,58$ for both the {weak} and {strong} interactions. The~effect of increasing the proton pairing strength is to partially smooth out these shell~effects.

\unskip

\subsubsection*{Energy Minimization with~ETFSI}\label{subsubsec:min}

A key ingredient of the ETFSI calculation is the determination of the seven parameters of the density profiles, using a minimization procedure.
For this case we have used the Python library {SciPy}~\cite{scipy1.0contributorsSciPyFundamentalAlgorithms2020}. For~a given $\rho_b$ and $Z$, a~initial guess is made for the number of neutrons, and~the corresponding $R_{WS}$ is calculated using Equation~\eqref{eqn:tot_particles}. The~parameters of the density profiles in Equation~\eqref{eqn:fd_profile} are varied, subject to constraints outlined in reference~\cite{onsi2008semi}, to~minimize the total energy of the WS cell (Equation~\eqref{eqn:etot}), which is calculated using a code written in Fortran~90. The~SI correction and proton pairing energy is then added perturbatively, which is necessary to prevent anomalously large values for the SI correction~\cite{pearsonInnerCrustNeutron2012}. This energy minimization is systematically repeated with different neutron numbers to find the cell configuration with the minimum~energy.

After repeating this process for every even value in $20\leq Z\leq60$, one finds the $Z$ that yields the lowest energy per particle~$e$; this is the optimum $Z$ for a given $\rho_b$.

Attempts to use the full 4th-order expressions for densities, along with the profiles using an extra damping factor as used in~\cite{onsi2008semi,pearsonInnerCrustNeutron2012,pearsonUnifiedEquationsState2018}, made the minimization much more difficult. Even without the damping factor, at~$\rho_b\approx0.05$~fm$^{-3}$ and above, our minimization procedure begins to fail increasingly often. This is the same difficulty reported by the authors of reference~\cite{pearsonInnerCrustNeutron2012}. Furthermore, the~presence of non-spherical clusters is expected above $\rho_b\approx0.05$~fm$^{-3}$~\cite{pearsonUnifiedEquationsState2020}. Since the goal of the present article is not to provide a complete EoS for calculations of an entire NS, we have limited our investigation to the range of baryonic densities $0.00025$~fm$^{-3}\leq\rho_b\leq0.048$~fm$^{-3}$.

\subsection{HFB vs. ETFSI+Pairing}

In this section we compare the results obtained in WS cells from solving the HFB equations with those from the ETFSI method, with~and without the additional correction for neutron pairing~correlations.

\begin{figure}[H]
\centering
\scalebox{.85}[.85]{\includegraphics{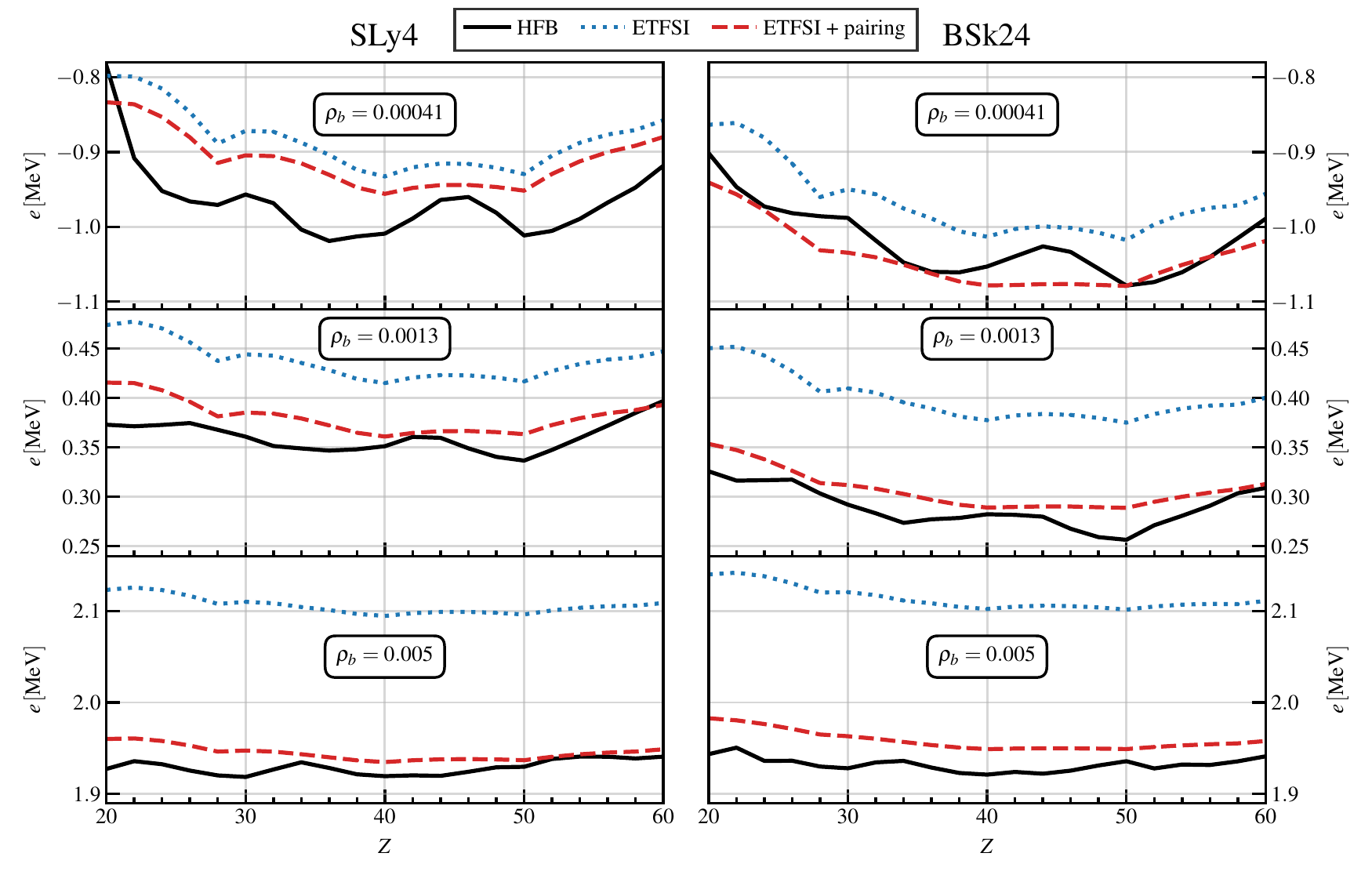}}
\caption{Selected slices of the energy surface at fixed baryonic densities $\rho_b$, showing the variation of $e$ with $Z$. Black solid lines show the HFB results, blue dotted lines the ETFSI results and the red dashed lines the results for ETFSI with neutron and proton pairing included. Results with the SLy4 functional (with {strong} pairing) are shown in the left column, and~those with BSk24 are shown in the right column. The~HFB calculations use the optimum values of $R_{WS}$ found with the ETFSI+pairing~method.}
\label{fig:etf_hfb_slices}
\end{figure}

In Figure~\ref{fig:etf_hfb_slices}, we show the results obtained with ETFSI, ETFSI+pairing and HFB, for~the WS cells at a few selected $\rho_b$, using the functionals SLy4 and BSk24. For~the SLy4 functional, the~{strong} pairing was used. The~results obtained with  BSk21 were almost identical to BSk24, so we do not show them in the figure.
For each WS cell, the~energy minimization is performed using the full ETFSI method, with~or without pairing. The~parameters of the resulting WS cells ($\rho_b, Z, R_{WS}$) were used to perform HFB calculations using the same functionals.
We have previously performed fully self-consistent HFB calculations~\cite{pastore2017new} using the SLy4 functional and the {strong} pairing interaction given in Equation~(\ref{eq:pair}), not~using the ETFSI cell parameters. We found that, while this leads to slightly different minima, they all still lie within the error bars estimated in that~work.

The energy dependence as a function of $Z$ for a full ETF calculation would be a smooth parabola, but~due to the Strutinsky integral correction, we clearly observe a modification to the total energy resulting from shell structure.
By comparing the ETFSI calculations with and without pairing, we~observe that the positions of the minima do not change, but~we observe a general reduction in the relative energy difference between the shell or sub-shell closure values and between the other WS cells. As~discussed in reference~\cite{pearson2015role}, this is mainly the effect of proton pairing. The~neutron pairing acts to globally shift the total energy as shown in Figure~\ref{fig:e_contributions}.

The HFB results are remarkably close to the ETFSI+pairing ones: near the drip density, \mbox{$\rho_b=0.00041$~fm$^{-3}$}, there is a larger discrepancy between HFB and ETFSI+pairing. This energy difference is of the order of $\approx$100~keV per particle. At~these low densities, the~energy from the cluster is greater than that from the neutron gas, and~so the different density profiles of the ETFSI and HFB methods, seen in Figure~\ref{fig:densities_fields}, are more important. The~correction for neutron pairing is also less accurate at these very low~densities.

For both SLy4 and BSk24, we find the two local minima at $Z=40$ and $Z=50$ for ETFSI+pairing, as~also found by other works~\cite{pearsonInnerCrustNeutron2012,pearson2015role,pearsonUnifiedEquationsState2018}. However, the~minima found with HFB take several values between $Z=36$ and $Z=50$.
The small discrepancy could be related to the role of neutron shell effects of the cluster that are not taken into account within the ETFSI method. From~$\rho_b=0.013$~fm$^{-3}$, when the neutron gas contribution starts to become more important, there is a remarkable agreement between the full HFB calculation and the ETFSI+pairing. This result confirms our previous hypothesis concerning the discrepancy observed between the two methods in reference~\cite{shelley2020accurately}.
At this density, the~total energy difference between the two methods is less than 10~keV per particle. The~energy minimum obtained with HFB is at $Z=50$, while the one with ETFSI+pairing is at $Z=40$. As~discussed in reference~\cite{pastore2017new}, the~accuracy of our HFB code is $\approx$4--5~keV per particle. Inspecting the figure, we~notice that the relative energy difference between the HFB configuration at $Z=40$ and $Z=50$ is $\approx$2~keV per particle, clearly within the error of our~calculations.

At higher baryonic density $\rho_b=0.005$~fm$^{-3}$ the agreement between the HFB and ETFSI+pairing remains, although~the energy minima obtained by the two calculations do not agree. However they are still compatible due to the error we estimate for the HFB calculations. From~this figure, we conclude that the inclusion of neutron pairing correlations in the ETFSI method leads to a very good agreement in the total energy per particle with the more involved HFB~calculation.

%%%%%%%%%%%%%%%%%%%%%%%%%%%%%%%%%%%%%%%%%%%%%%%%%%%%%%%%%%%%%%%%%%%%%%%%%%%%%%%%%%%%%%%%%%%%%%%%%%%%

\section{Results}\label{sec:results}

% Having demonstrated the reliability of the ETFSI+pairing method, it is now possible to use it for systematic calculations of the inner crust.
Having demonstrated the compatibility between the ETFSI+pairing and HFB methods, it is now possible to use ETFSI+pairing for systematic calculations of the inner crust.
We first tested our method, without~neutron pairing, against~previous work. We used BSk21 and SLy4, as~in reference~\cite{pearsonInnerCrustNeutron2012}, and~BSk24, as~in reference~\cite{pearsonUnifiedEquationsState2018}. For~all of these functionals, we found that the energy minimum occurs at $Z=40$ for all densities in the range $0.00025$~fm$^{-3}\leq\rho_b\leq0.048$~fm$^{-3}$ (as in the previous work), except~for at very low densities, less than $\approx$0.01~fm$^{-3}$, where we find $Z=50$. As~noted in~\cite{pearsonInnerCrustNeutron2012}, the~two local minima corresponding to $Z=40$ and $50$ are very close in energy at low densities. This~result is independent on the use or not of the energy correction given by neutron pairing~correlations.

We attribute the discrepancy in our results to our treatment of the 4th-order ETF contributions to the energy (explained in Section~\ref{subsec:etf}). The~smaller clusters found at these lower densities have less diffuse surfaces. Therefore, using the full 4th-order expressions with its higher-order derivatives is more vulnerable to numerical inaccuracies. A~possible source of error is the way the total energy is calculated. These small numerical discrepancies are related to the method of numerical integration and we have seen that they are sufficient to explain the different minima observed at very low density, as~discussed in reference~\cite{pearsonInnerCrustNeutron2012}.

\begin{figure}[H]
\centering
\includegraphics{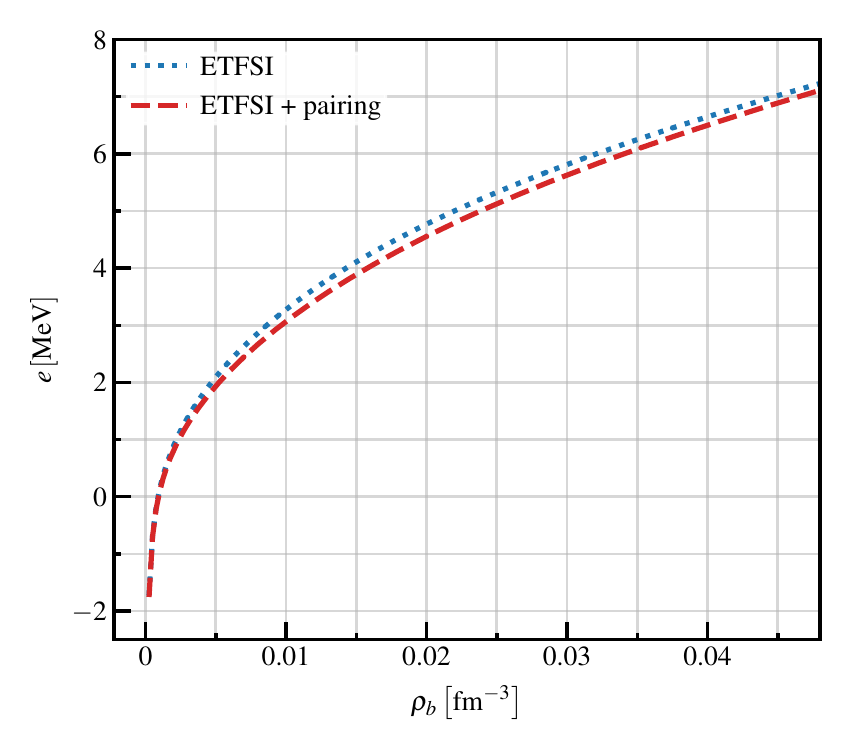}
\caption{Energy per particle, $e$, for~the inner crust, using the SLy4 functional. The~result for ETFSI is shown by the blue dotted line, and~the red dashed line shows the result with neutron and proton pairing included (using the {strong} pairing).}
\label{fig:e_rho}
\end{figure}

In Figure~\ref{fig:e_rho}, we present the equation of state obtained with a full ETFSI minimization, using the SLy4. ETFSI results are shown with blue dotted lines, and~ETFSI+pairing results using the {strong} pairing interaction with red dashed lines. As~expected, the~inclusion of neutron pairing effects in the system decreases the energy per particle of the system, as~shown in Figure~\ref{fig:e_rho}, but~it does not affect the global~trend.

\begin{figure}[H]
\centering
\scalebox{.85}[.85]{\includegraphics{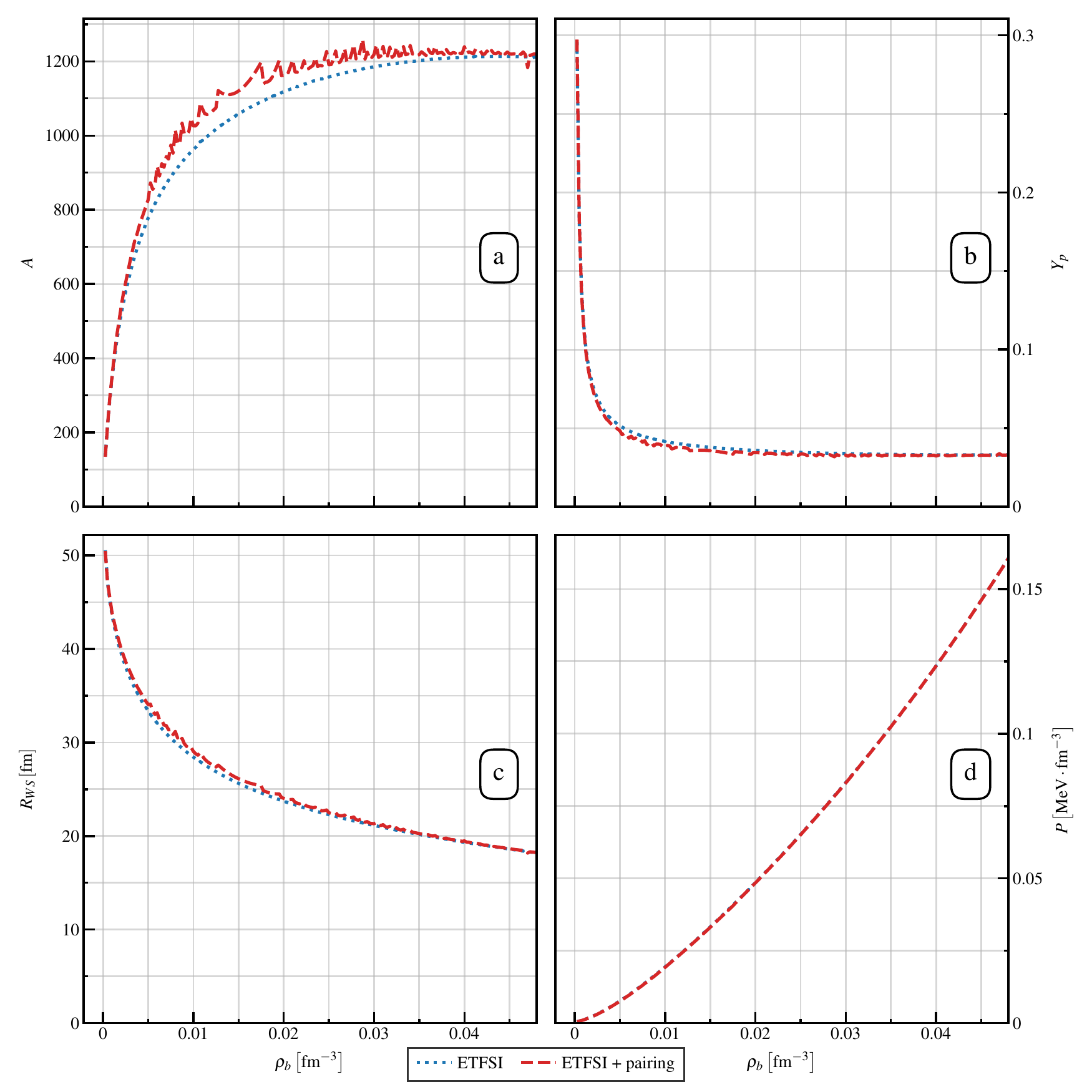}}
\caption{The variation through the inner crust of: the total particle number $A$ (panel \emph{a}), proton fraction $Y_p$ (panel \emph{b}), Wigner--Seitz cell radius $R_{WS}$ (panel \emph{c}) and pressure $P$ (panel \emph{d}). ETFSI results are shown with blue dotted lines, and~those for ETFSI+pairing are shown with red dashed~lines.}
\label{fig:globalProperties_rho}
\end{figure}

In Figure~\ref{fig:globalProperties_rho}, we compare other properties of the WS cells obtained using SLy4 with and without pairing. In~panel~\emph{a} of Figure~\ref{fig:globalProperties_rho}, we observe a larger total nucleon number $A$ is the case of ETFSI+pairing. This reflects in a small reduction of the proton fraction~$Y_p=Z/A$ as shown in panel~\emph{b}, and~a small increase of the WS cell radius~$R_{WS}$, shown in panel~\emph{c}. This adds to the reliability of the ETFSI method, whose semiclassical Wigner--Kirkwood expansion is exact in the limit of PNM ($Y_p\to 0$).

The pressure, defined as
\begin{eqnarray}\label{eqn:pressure}
P=-\left(\cfrac{\partial E}{\partial V}\right)_{T,N,Z}\;,
\end{eqnarray}
is calculated following closely the approach described in Appendix~B of~\cite{pearsonInnerCrustNeutron2012}. In~the presence of neutron pairing, the~pressure of the cell decreases, but~as panel~\emph{d} shows this difference is negligible, certainly~far smaller than the difference that would arise from using different~functionals.

All of these observed differences are at their greatest around $\rho_b=0.015~$fm$^{-3}$. %Shelley: 0.15 changed to 0.015, following comment from reviewer
This is near the densities where the PNM pairing gaps, shown in Figure~\ref{fig:PNM_gaps}, are at their maximum. The~small bumps for the ETFSI+pairing case in Figure~\ref{fig:globalProperties_rho}, most notably in panel \emph{a}, can be attributed to the inclusion of the SI correction and pairing in the second minimization step, when the total particle number~$A$ is determined (see Section~\ref{subsubsec:min}).

We find that, for~all functionals and densities tested, the~inclusion of neutron pairing does not change the optimum $Z$. This is easily understood by looking at the upper-right panel of Figure~\ref{fig:e_contributions}, where the contribution to $e$ from neutron pairing is almost constant with respect to $Z$.
%%%%%%%%%%%%%%%%%%%%%%%%%%%%%%%%%%%%%%%%%%%%%%%%%%%%%%%%%%%%%%%%%%%%%%%%%%%%%%%%%%%%%%%%%%%%%%%%%%%%

\section{Conclusions}\label{sec:concl}

We have presented a systematic comparison between solving the HFB equations and using the ETFSI method, using exactly the same numerical conditions. We have observed that the inclusion of neutron pairing correlation in ETFSI using a simple LDA approximation leads to a remarkable reduction in the discrepancy between the two methods, thereby confirming the success of using ETFSI+pairing for the determination of the properties of the WS in the inner~crust.

After neutron pairing is included in the modeling of the inner crust, we find no change in the prediction of the optimum proton number~$Z$ at a given baryonic density~$\rho_b$, for~any of the functionals SLy4, BSk21 and BSk24. The~location of the minimum of the nuclear contribution, dictated by the choice of functional, has a much bigger influence on the optimum $Z$.

The additional neutron pairing energy contribution leads to a general energy reduction in the Wigner--Seitz  cell across the range $20\leq Z\leq60$ investigated. The~most interesting effect of neutron pairing is the increase in the radius of the WS cells and the number of neutrons per cell, which means a small decrease in the proton fraction is observed. The~pressure is also slightly decreased, but~the change is not~significant.

We conclude that ETFSI+pairing method is capable of giving a very accurate description of the structure of the inner crust. The~results obtained are of the same quality as more advanced HFB calculations done under the same numerical conditions. We recall that given the proximity in energy of the minima, small numerical inaccuracies may lead to different minima, thereby explaining the apparent discrepancy of results within the scientific~literature.

The reliability of the ETFSI+pairing method calls for a {hybrid} approach to inner crust calculations: HFB would be still be used at lower densities, where it has a superior quality, and~at higher densities up to the crust-core transition, ETFSI+pairing would be~used.

%%%%%%%%%%%%%%%%%%%%%%%%%%%%%%%%%%%%%%%%%%%%%%%%%%%%%%%%%%%%%%%%%%%%%%%%%%%%%%%%%%%%%%%%%%%%%%%%%%%%
\vspace{6pt}

\authorcontributions{Both authors contributed equally to the investigation and to the writing of the article. All~authors have read and agreed to the published version of the manuscript.}%Shelley: Added
%Please add 'Author Contributions: For research articles with several authors, a short paragraph specifying their individual contributions must be provided.

\funding{This work was supported by STFC Grant No. ST/P003885/1.} %Shelley: Funding added
% Please disclose any funding informa, or add "This research received no external funding.".

%\section*{Acknowledgments}

\acknowledgments{We thank Michael Urban for useful discussions around neutron superfluidity. We also thanks M. Pearson, N. Chamel and S. Goriely for helping us in debugging the ETFSI numerical~code.} %Shelley: Funding removed from acknowledgments

%%%%%%%%%%%%%%%%%%%%%%%%%%%%%%%%%%%%%%%%%%%%%%%%%%%%%%%%%%%%%%%%%%%%%%%%%%%%%%%%%%%%%%%%%%%%%%%%%%%%
\conflictsofinterest{The authors declare no conflict of interest.} %Shelley: Added
%Please disclose any conflicts of interest, or add ``The authors declare no conflicts of interest''.

%%Declare conflicts of interest or state ``The authors declare no conflict of interest.'' Authors must identify and declare any personal circumstances or interest that may be perceived as inappropriately influencing the representation or interpretation of reported research results. Any role of the funders in the design of the study; in the collection, analyses or interpretation of data; in the writing of the manuscript, or in the decision to publish the results must be declared in this section. If there is no role, please state ``The funders had no role in the design of the study; in the collection, analyses, or interpretation of data; in the writing of the manuscript, or in the decision to publish the results Appendix \ref{app}''.
%%Please add it.
%\begin{appendix}

%\section{Neutron Condensation Energy}\label{app:cond}
\appendixtitles{yes} % Leave argument "no" if all appendix headings stay EMPTY (then no dot is printed after "Appendix A"). If~the appendix sections contain a heading then change the argument to "yes".
\appendix
\section{Neutron Condensation~Energy}\label{app:cond}

We derive here the expression for the pairing condensation energy per particle for PNM, given in Equation~\eqref{eqn:lda}. For~simplicity we set $\hbar=m=1$.
Since we use a contact pairing interaction, the~pairing gap in PNM is momentum-independent. The~single-particle energy, relative to the effective chemical potential, is given by
\begin{eqnarray}\label{eqn:spke}
\xi_k = \frac{k^2}{2}+U-\mu\;,
\end{eqnarray}
where $\mu$ is the effective chemical potential, i.e.,~scaled respect to the HF mean field, and~$k$ is the particle momentum. The~quasi-particle energy is then given by
\begin{eqnarray}\label{eqn:qpe}
E_k = \sqrt{\xi_k^2 + \Delta^2}\;,
\end{eqnarray}
where $\Delta$ is the pairing gap. We write the energy per unit volume of a superfluid system as
\begin{eqnarray}\label{eqn:superf_edens}
E(\mu,\Delta) = \int_0^\Lambda\frac{k^2}{2\pi^2}\left[\frac{-\Delta^2}{2E_k} + \frac{k^2}{2}\left(1-\frac{\xi_k}{E_k}\right)\right]dk\;,
\end{eqnarray}
where $\Lambda$ is the cut-off momentum. After~integration, the~first term of the integrand of Equation~\eqref{eqn:superf_edens} yields the pairing energy density, and~the second term the kinetic energy density corrected by the depletion of the occupation factors~\cite{ring2004nuclear}. The~kinetic energy density per unit volume of a non-superfluid system with the same density is
\begin{eqnarray}\label{eqn:nonsuperf_edens}
E_0 = \frac{k_F^5}{10\pi^2}\;.
\end{eqnarray}

In a superfluid system with a given $\mu$, the~density can be calculated as
\begin{eqnarray}\label{eqn:dens}
\rho(\mu,\Delta) = \int_0^\Lambda\frac{k^2}{2\pi^2}\left(1-\frac{\xi_k}{E_k}\right)dk\;.
\end{eqnarray}

We now estimate the true energy gain per particle, in~terms of the quantities expressed in Equations~\eqref{eqn:superf_edens}--\eqref{eqn:dens} as
\begin{eqnarray}\label{eqn:delta_e}
\delta\epsilon = \left(\frac{E-E_0}{\rho}\right)\;.
\end{eqnarray}

The previous equation can be simplified in the weak coupling limit, where $\Delta\ll\mu$. In~this case, the~change in the kinetic energy density and in the density of the system is negligible, so Equation~\eqref{eqn:delta_e} reduces to
\begin{eqnarray}\label{eqn:delta_e_weak}
\delta\epsilon_{weak} = -\frac{3\Delta^2}{8\mu}\;.
\end{eqnarray}

This expression is in agreement with the one given in reference~\cite{PhysRevLett793347}.
Equation~\eqref{eqn:delta_e_weak} can be further simplified by approximating the chemical potential $\mu$ with the Fermi energy $e_F=(3\pi^2\rho_n)^{1/3}$.

\begin{figure}[H]\label{fig:weak_coupling}
\centering
\includegraphics{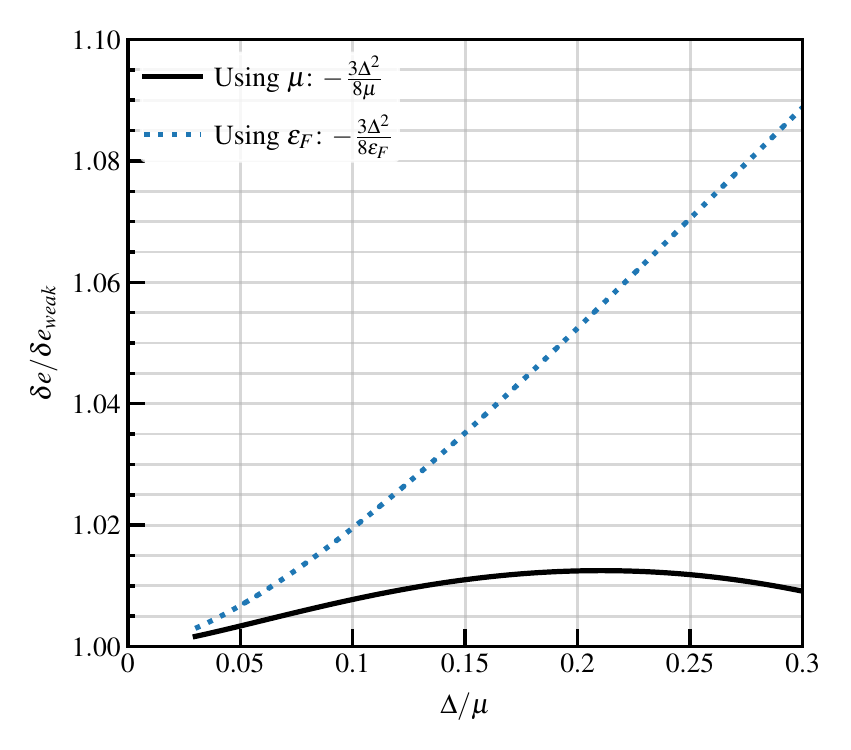}
\caption{Solid black line shows the ratio of the weak-coupling approximation for $\delta\epsilon$ (Equation~\eqref{eqn:delta_e_weak}) to the exact result (Equation~\eqref{eqn:delta_e}). The~blue dotted line shows the same, but~with $\mu$ in Equation~\eqref{eqn:delta_e_weak} replaced by $\varepsilon_F$.}
\label{fig:LDA_accuracy}
\end{figure}

In Figure~\ref{fig:LDA_accuracy}, we compare the validity of the weak coupling limit by comparing the result exact result given by Equation~\eqref{eqn:delta_e} and the two different approximations. It shows the ratio of Equation~\eqref{eqn:delta_e} to Equation~\eqref{eqn:delta_e_weak}, using either $\mu$ or $\varepsilon_F$ in the denominator of Equation~\eqref{eqn:delta_e_weak}. We vary the ratio $\Delta/\mu$ over the range~0.03--0.3.

We observe that the weak coupling limit gives a very nice reproduction of the total energy correction with an error of $\approx$1\% over relevant range of variation.
In the regions of the star where $\Delta\ll\mu$, the~use of $\varepsilon_F$ instead of $\mu$ is then fully justified. The~weak coupling approximation with $\varepsilon_F$ tends to give a larger error especially in the low-density region of the star, but~such an error is still less than 10\%, and~is probably less important than other approximations in the ETFSI method~\cite{carreau2019bayesian}.
%\end{appendix}

%%%%%%%%%%%%%%%%%%%%%%%%%%%%%%%%%%%%%%%%%%%%%%%%%%%%%%%%%%%%%%%%%%%%%%%%%%%%%%%%%%%%%%%%%%%%%%%%%%%%

%\bibliography{bib}
\reftitle{References}
\bibliographystyle{Definitions/mdpi}

\end{document}